\documentclass[aps,twocolumn,showpacs,superscriptaddress]{revtex4-1}
\usepackage{amssymb}
\usepackage{amsmath}
\usepackage{graphicx}
\usepackage{hyperref} 
\hypersetup{colorlinks=true,
            linkcolor=blue,
            citecolor=blue,
            urlcolor=orange}
        
\usepackage[resetlabels,labeled]{multibib}
\newcites{App}{App Readings}
\usepackage[dvipsnames]{xcolor}

\begin{document}

\title{
Local magnetic moment formation and Kondo screening \\in the {half-filled} single-band Hubbard model}
\author{T. B. Mazitov}
\affiliation{Center for Photonics and 2D Materials, Moscow Institute of Physics and Technology, Institutsky lane 9, Dolgoprudny,
141700, Moscow region, Russia}
\author{A. A. Katanin}
\affiliation{Center for Photonics and 2D Materials, Moscow Institute of Physics and Technology, Institutsky lane 9, Dolgoprudny,
141700, Moscow region, Russia}
\affiliation{M. N. Mikheev Institute of Metal Physics, Kovalevskaya Street 18, 620219
Ekaterinburg, Russia.}

\begin{abstract}
We study {the} formation of local magnetic moments in {the} strongly correlated Hubbard model within dynamical {mean-field} theory and associate peculiarities of {the} temperature dependence of local charge $\chi_c$ and spin $\chi_s$ susceptibilities with different stages of local moment formation. {The} local maximum of {the} temperature dependence of the charge susceptibility $\chi_c$ is associated with {the} beginning of local magnetic moment formation, while the minimum of the susceptibility $\chi_c$ and double occupation, as well as the {low-temperature} boundary of the plateau of the effective local magnetic moment $\mu_{\rm eff}^2=T\chi_s$ temperature dependence
are connected with {the} full formation of local moments. We also obtain the interaction dependence of the Kondo temperature $T_K$, which is compared to the fingerprint criterion of Chalupa \textit{et al.} [Phys.  Rev.  Lett. {\bf 126},  056403 (2021)]. Near the Mott transition the two criteria coincide, while further away from the Mott transition the fingerprint criterion somewhat overestimates the Kondo temperature.  The relation of the observed features to the behavior of eigenvectors/eigenvalues of fermionic frequency-resolved charge susceptibility and divergences of irreducible vertices is discussed.   

\end{abstract}

\maketitle



The localization of electrons in solids by correlation (interaction) effects yields the formation of local magnetic moments, which are crucial for explaining the observable magnetic properties of some of the existing materials and predicting new magnetic materials.
Typical examples of {the} importance of local magnetic moments include some aspects of the physical properties of {high-temperature} superconductors in the underdoped regime \cite{Lee,Sushkov}, modern explanations of the ferromagnetism of transition metals
(see, e.g., Refs. \cite{OurIron,OurFeNi,Hausoel,OurFe1,OurFeGamma}), as well as the magnetic properties of iron pnictide superconductors \cite{Pnictides,Pnictides1}.
Local magnetic moments in the {above-mentioned} substances appear due to electronic correlations in proximity to the (orbital-selective) interaction-induced Mott {metal-insulator} transition (MIT) (see, e.g., the discussion in Refs. \cite{Spalek,Nozieres,Fabrizio}), and/or due to the Hund's exchange interaction \cite{Pnictides,Pnictides1,ThreeOrbital,OurFe1,Delft}.  

 Although the concept of MIT was introduced by Mott in 1949 \cite{Mott}, quantitative studies of the Mott transition became possible with the discovery of the dynamical mean-field theory (DMFT) \cite{DMFT}. Originally, the MIT
 was described mainly on the basis of single-particle properties, e.g., spectral functions, densities of states, etc. The three-peak structure of the density of states near MIT
 reflects {the} coexistence of localized electrons (corresponding to the states in the Hubbard subbands) with itinerant degrees of freedom, described by the quasiparticle peak (see, e.g., Refs. \cite{DMFT,Fabrizio}).
  The developments of the nonlocal diagrammatic extensions of DMFT \cite{OurRev} yielded new insight on the nonperturbative aspects of MIT
  via studying the divergences of the two-particle irreducible vertices \cite{ToschiDiv1,Schafer,ToschiCompress,ToschiAnd1,ToschiInterplay,ToschiDoping,AtomicHub}. These divergences were interpreted as precursors of local moment formation
  \cite{ToschiAnd1,Toschi}. The formation of local magnetic moments was also recently discussed within the nonlocal extensions of DMFT in Ref. \cite{Katsnelson}.

In the presence of conduction (itinerant) electrons (i.e., on the metallic side of MIT)
the local moments are screened below a certain characteristic (Kondo) temperature. In contrast to the standard Kondo effect, in strongly correlated substances 
the role of magnetic impurities is played by naturally occurring local magnetic moments and 
the same electrons participate in {the} formation of local moments and their screening. This reflects the dual role of $d$ electrons, which was first discussed for transition metals by  Vonsovskii \cite{Vonsovskii} and more recently emphasized for pnictides \cite{dual1,dual2,dual3,dual4}.  
Although the presence of a characteristic (Kondo) temperature $T_K$ near MIT,
below which almost formed local moments are screened by itinerant electrons, was emphasized in the early stages of DMFT studies \cite{Pruschke} {and its relation to the frequency dependence of the electronic self-energy and spectral functions was discussed \cite{Pruschke,Bulla,HeldScal}}, the properties of Kondo screening near 
MIT were not intensively studied. 
{Being generally larger than the Fermi liquid coherence temperature \cite{Coh}, 
$T_K$ determines at the same time the spin dynamics at a given lattice site, 
which makes this temperature physically important}.

The Kondo temperature of local magnetic moments in strongly correlated systems can be extracted from {a} comparison of the local spin susceptibility to that for the Kondo model \cite{Wilson,Wilson1}. 
This approach was applied to extract the Kondo temperature of Hund's metals \cite{Hausoel,Hund1,Hund2,Hund3,OurFeNi,Comment,Reply}, as well as for the description of Kondo screening in the Anderson impurity model \cite{Toschi,ToschiAnd1} and the Hubbard model in the vicinity of 
MIT \cite{Pruschke,Comment}. Therefore, it provides a unified view on the Kondo screening in strongly correlated substances. 

We note that while the single-impurity Kondo model can be considered as an effective low-energy model for the Anderson impurity model, its applicability for describing screening in 
lattice models, such as the Hubbard model, is not \textit{a priori} clear. On the other hand, due to {the} reduction of the lattice problem to the impurity problem by DMFT, one can hope that at least within this theory the Kondo model is an appropriate effective low-energy model for lattice problems too. 

Recently, the two-particle criterion for the Kondo temperature in terms of frequency-dependent charge susceptibility was formulated for the Anderson impurity model in Ref. \cite{Toschi}. It was suggested that this criterion also applies to the Hubbard model in the vicinity of MIT.
The 
generalization of this criterion for multiorbital systems and fillings away from half filling is however not obvious. 
{A} somewhat different criterion of local moment formation was also proposed in Ref. \cite{Katsnelson}.


In the present Letter we consider the formation of local magnetic moments in {a} single-band strongly correlated system and their screening properties on the verge of MIT.
We study local charge susceptibilities and spin susceptibilities within DMFT to provide a unified view of local magnetic moment formation in the model considered.

In particular, we address the following topics: 
  (i) the interaction dependence of the temperatures of the beginning and the full formation of local magnetic moments, as well as their screening (Kondo) temperature, and  
  (ii) the connection of Kondo screening to peculiarities of static charge susceptibility and double occupancy. 


{\it Model and method}. We consider a half-filled Hubbard model on the square lattice ({the obtained results are however expected to be qualitatively applicable for an arbitrary density of states})
\begin{equation} \label{eq:hubbard}
H = - t\sum_{\langle i, j \rangle , \sigma}{c_{i\sigma}^\dagger c_{j\sigma}} + U\sum_i{n_{i\uparrow}n_{i\downarrow}},
\end{equation}
and use the half bandwidth $D=4t=1$ as the unit of energy. 

Due to the assumption of locality of the self-energy, the DMFT \cite{DMFT}
is a convenient tool to study the formation and screening of local magnetic moments, which can be directly traced at the impurity site. To trace the formation of local moments we calculate in the self-consistent solution of DMFT the local spin susceptibility 
$\chi_s(i\omega_n) = \int_0^\beta \left\langle S^z(\tau) S^z(0) \right\rangle \exp(i\omega_n\tau)d\tau,$
where $S^z(\tau)$ is the impurity spin projection at the imaginary time $\tau$, $\beta=1/T$ (Boltzmann's constant is put to unity), and $\omega_n=2n\pi T$ are the bosonic Matsubara frequencies. We also consider 
local static charge susceptibility (local charge compressibility) $dn/d\mu=\chi_c(T)$, where the change of the chemical potential $d\mu$ acts only at the impurity site,
\begin{equation} 
\chi_c(T)=\int_0^\beta \left( \left\langle n(\tau) n(0) \right\rangle - \left\langle n(0) \right\rangle^2 \right) d \tau=\sum_{\nu\nu'}\chi_c^{\nu\nu'},\label{ChargeCorr}
\end{equation} 
$n(\tau) = \sum_\sigma c_{i\sigma}^\dagger(\tau) c_{i\sigma}(\tau)$,  and Matsubara fermionic frequency $\nu,\nu'$-resolved susceptibilities $\chi_c^{\nu\nu'}$ are expressed via 
two- and single-particle impurity Green's functions (see Supplemental Material \cite{SM}).

For computations, we {mainly} use the {continuous-time} quantum Monte Carlo (CT-QMC) impurity solver, implemented in the 
iQIST software package \cite{iQIST,iQISTNote}. At strong coupling ($U\geq 2.3$) near MIT
we use numerical renormalization group (NRG) approach \cite{NRG} within TRIQS-NRG Ljubljana interface package \cite{TRIQS}. 


\begin{figure}[t]
\includegraphics[width=1.0\linewidth]{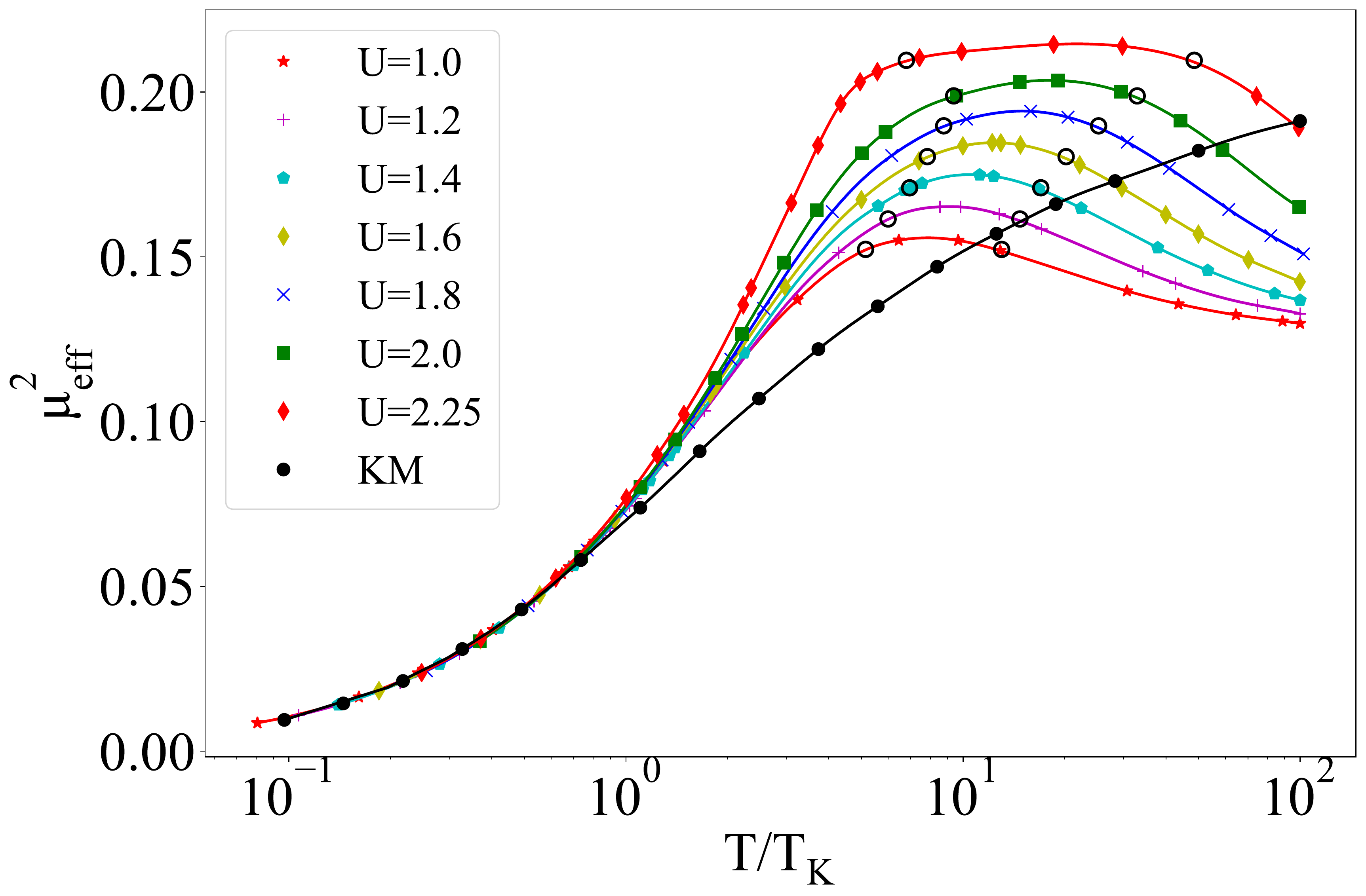}
\caption{(Color online). Temperature dependence of the square of the effective local moment $\mu_{\rm eff}^2=T\chi_s(0)$ at various values of the Coulomb interaction $U$. The Kondo temperature $T_K$ is obtained from the fit to the universal dependence for the Kondo model (KM) \cite{Wilson,Wilson1} (black line) at low temperatures. The open black circles denote 
the characteristic boundaries of the “plateau” of $\mu_{\rm eff}^2$, which is defined by the values of temperature at which $\mu_{\rm eff}^2$ reaches 0.975 of its maximal value.}
\label{fig:spin_susceptibility}
\end{figure}

 \begin{figure}[b]
    \includegraphics[width=0.95\linewidth]{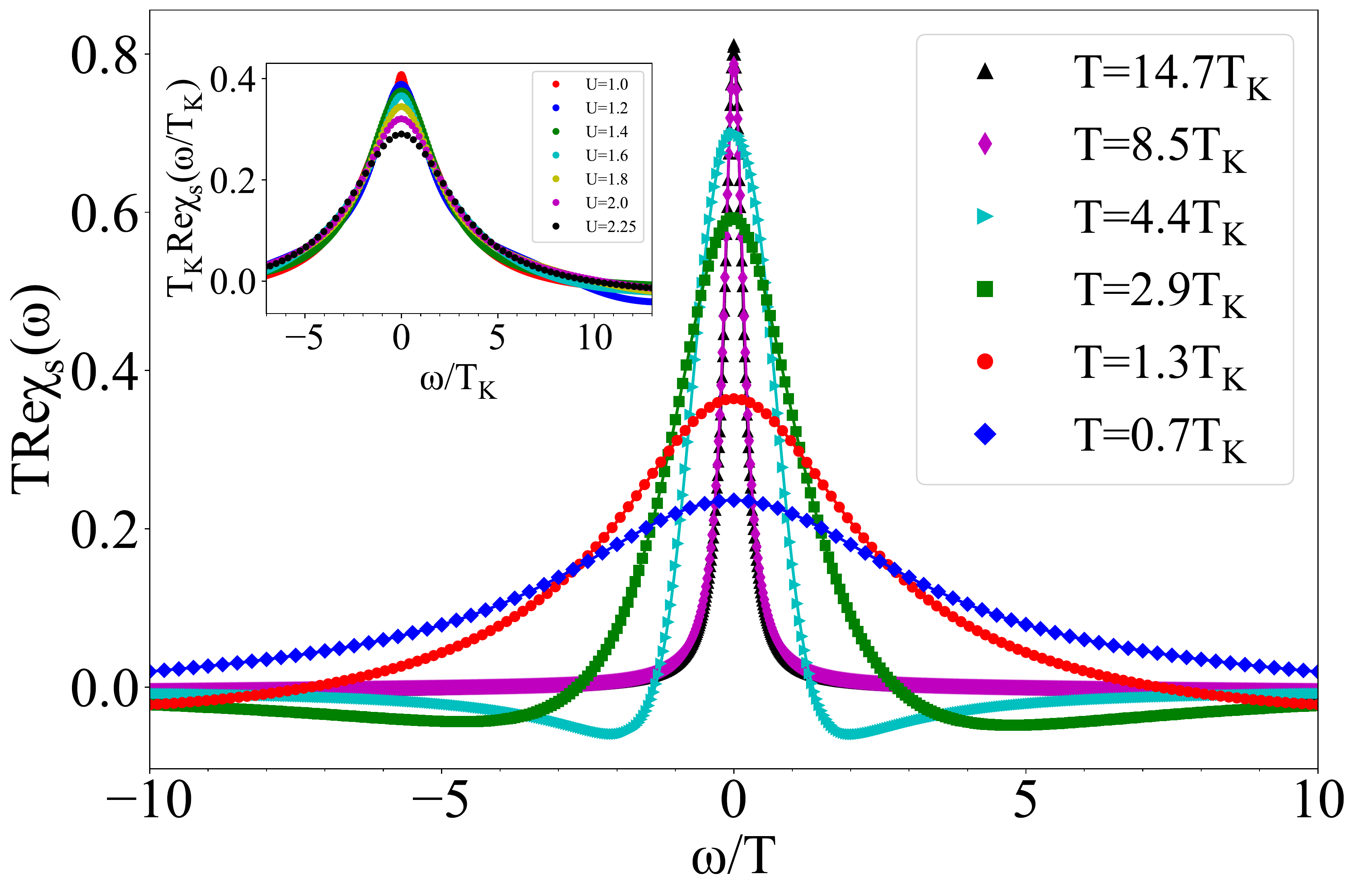}
    \caption{(Color online). Real frequency dependence of the real part of local spin susceptibility $\chi_s(\omega)$ at the value of the Coulomb interaction $U = 2$. The inset shows the frequency dependence in units of $T_K$ at various $U$ and $T=0.01$.}
    \label{fig:frequency_dependence}
\end{figure}

{\it Results}. We consider first the static local spin susceptibility $\chi_s(0)$ (see Fig. \ref{fig:spin_susceptibility}). To compare the obtained results with the Kondo model \cite{Wilson,Wilson1} and unambiguously determine the Kondo temperature, we plot the square of the effective local moment $\mu_{\rm eff}^2=T\chi_s(0)$ vs $T/T_K$, where $T_K$ is determined by the fit of low-temperature data to the results of the Kondo model (cf. Refs. \cite{ToschiAnd1,Toschi}). 
With increasing $U$ the maximum of the temperature dependence of $\mu_{\rm eff}^2$ forms a plateau at $5T_K\lesssim T\lesssim 50T_K$, whose height approaches $\mu_{\rm eff}^2= 1/4$, reflecting {the} formation of local magnetic moments. 
As the temperature is lowered, the effective local moment $\mu_{\rm eff}$ decreases due to screening by itinerant electrons. 
At $T\lesssim T_K$ the obtained $\mu_{\rm eff}^2$ approaches the universal temperature dependence for the Kondo model, which shows {the} complete screening of local moments in this temperature regime and the correctness of the definition of the Kondo temperature $T_K$.

In Fig. \ref{fig:frequency_dependence} we show the frequency dependence of local dynamic spin susceptibility $\chi_{s}(\omega)$
 on the real frequency axis (obtained by using Pade approximants \cite{Pade}).  
 The frequency dependence of the real part of susceptibility has a form of the peak, whose width reflects {an} inverse lifetime of local moments $\hbar/t_{\rm loc}$ \cite{OurFeGamma,Notew,Notetloc}. At $U=2$ in the temperature interval on the plateau of $\mu_{\rm eff}^2$ ($T\sim 10T_K$), the lifetime $t_{\rm loc}\sim\hbar/T$ shows well formed local moments. With a further decrease of temperature (screening regime) the 
 peak is strongly broadened ($Tt_{\rm loc}$ decreases) due to screening effects. At low temperatures we find almost universal frequency dependence with $t_{\rm loc}\sim \hbar/T_K$ (cf. Ref. \cite{HeldScal}).

\begin{figure}[t]
\centerline{\includegraphics[width=1.0\linewidth]{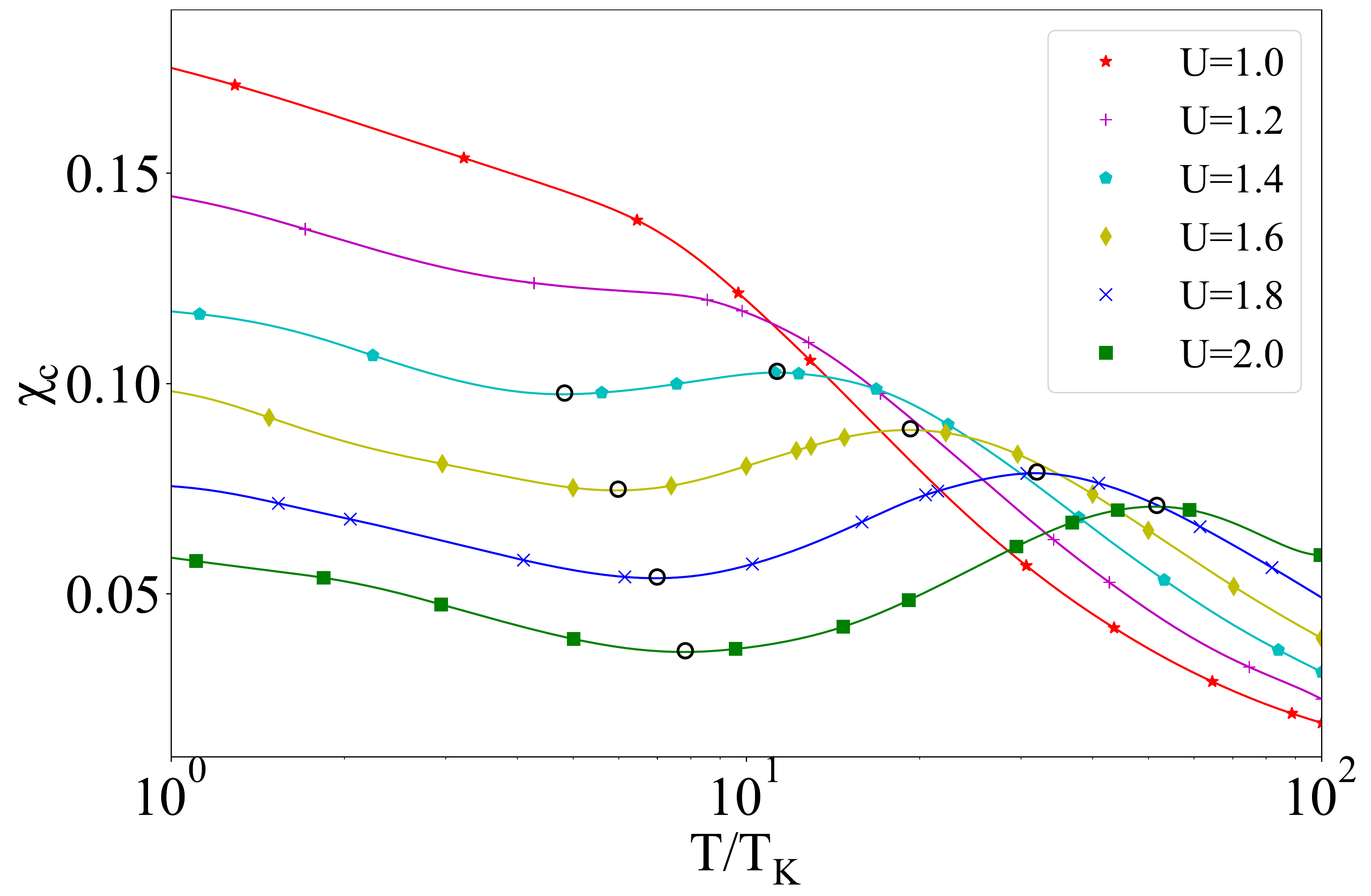}}
\caption{(Color online). Temperature dependence of local static charge susceptibility $\chi_c$ at various values of the Coulomb interaction $U$. The open black circles indicate local minima and maxima of $\chi_c$.}
\label{fig:charge_susceptibility}
\end{figure}

To study {the} behavior of charge degrees of freedom in the local moment and screening regimes, in Fig. \ref{fig:charge_susceptibility} we show {the} temperature dependence of local charge compressibility $\chi_c(T)$. 
With decreasing temperature the local compressibility first increases due to {an} increase of the coherence of quasiparticles. At lower temperatures, the decrease of local compressibility is observed, which is associated with local moment formation (cf. Refs. \cite{ToschiCompress,Toschi}). Therefore, the position of the maximum of $\chi_c(T)$ dependence, which occurs at $T_{c,{\rm max}}\sim(10$--$50)T_K$, is used in the following as a characteristic temperature of entering the preformed local moment (PLM) regime.
With further reducing temperature, at $T_{c,{\rm min}}\sim(5$--$10)T_K$ we observe {a} characteristic minimum of local compressibility, which we associate with the full formation of local moments, i.e.,
maximal portion of electrons participating in the local moment formation. A further increase of local compressibility reflects the screening of local moments (which is denoted in the following as the SCR regime), occurring as a consequence of virtual transitions from the local moment to itinerant states.

As we discuss in the Supplemental Material \cite{SM}, 
the increase of the local compressibility below $T_{c,{\rm min}}$ is provided by the lowest (negative) eigenvalues of susceptibility $\chi_c^{\nu\nu'}$ (corresponding to even in frequency eigenfunctions), which are related to the irreducible vertex divergences. 
We also compare \cite{SM} the above discussed temperature dependence of local compressibility to that for double occupation $\langle n_\uparrow n_\downarrow \rangle$, which describes the average value of the square of the local spin $\langle{\bf S}^2\rangle=({3}/{4})(1 - 2\langle{n_\uparrow n_\downarrow}\rangle)$. Similarly to local compressibility, the double occupation has a minimum at approximately the same temperatures $T_{c,{\rm min}}$. Notably, the double occupation only slightly increases below $T_{c,{\rm min}}$, in contrast to the local compressibility $\chi_c$, which almost recovers at low temperatures its maximal value at the temperature $T_{c,{\rm max}}$. This reflects the difference between electrons participating in virtual transitions and {the} number of electrons {participating} in screening {at a given time}. {Instantaneously}, only {a} small portion of electrons can participate in screening at half filling, since most of them already form local magnetic moments. However, due to virtual transitions, substantial screening effects can be achieved at a given site of the lattice {over long time scales}. According to the thermodynamic relation $(\partial S/\partial U)_T=-(\partial \langle n_\uparrow n_\downarrow \rangle/\partial T)_U$ (cf. Ref. \cite{triangular}), the entropy $S$ reaches a local maximum as a function of $U$ at the boundary of the PLM and SCR regions. {This reflects maximal spin degeneracy, which occurs in the regime of fully formed local moments.}

\begin{figure}[t]
\centerline{\includegraphics[width=1.0\linewidth]{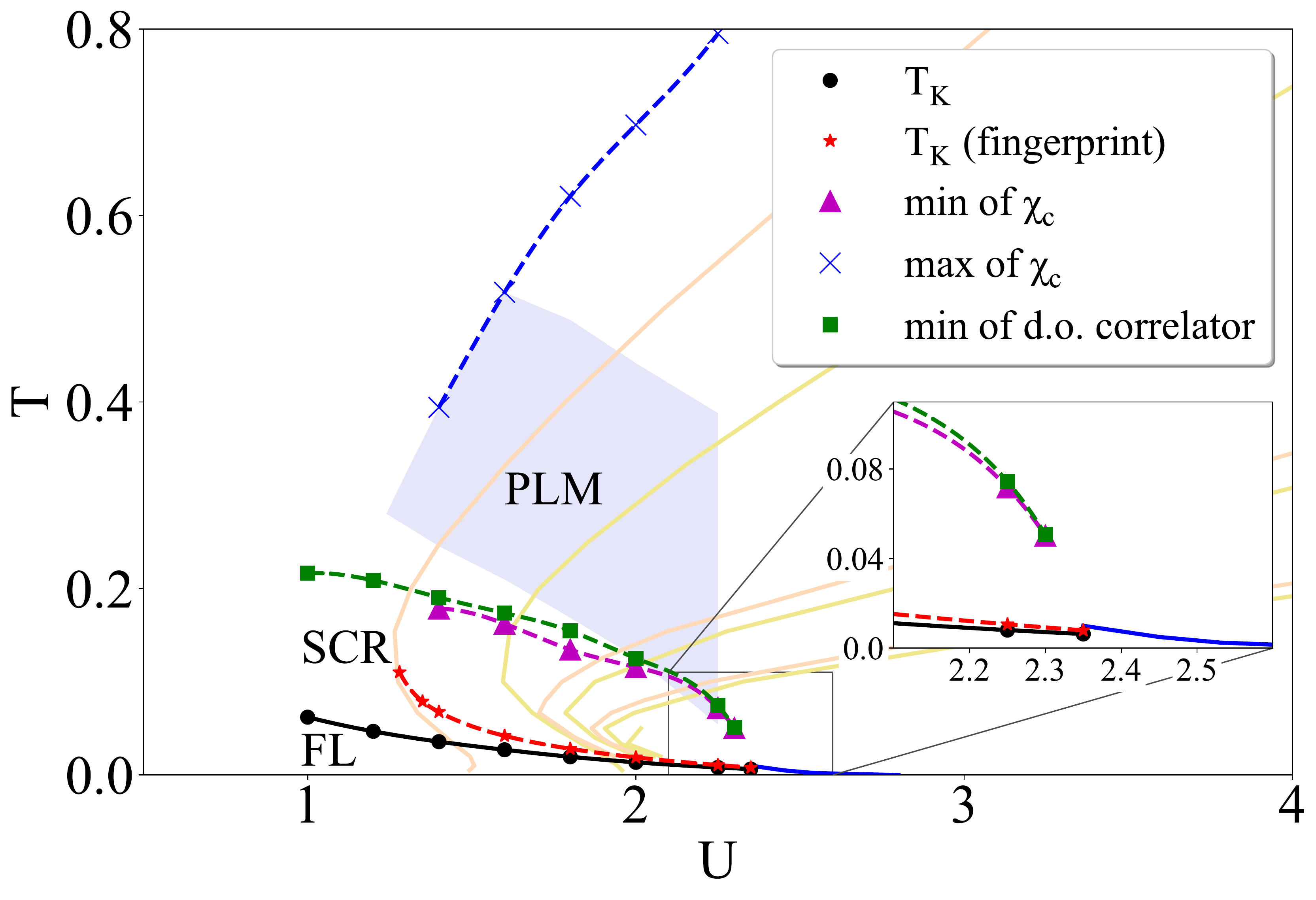}}
\caption{(Color online). Phase diagram showing the dependence on the Coulomb interaction $U$ of the Kondo temperature $T_K$ (black line with circles), the temperatures $T_{c,{\rm max}}$ and $T_{c,{\rm min}}$ of the maxima and minima of local charge compressibility $\chi_c(T)$ 
(blue dashed line with crosses and purple dashed line with triangles, respectively), and minima of double occupation 
(green dashed line with squares).
The shaded area corresponds to the ``plateau'' of $\mu_{\rm eff}^2(T)$ from Fig. \ref{fig:spin_susceptibility}, bounded by the temperatures $T_{c,{\rm max}}$. The red dashed line with asterisks shows the Kondo temperature according to the ``fingerprint'' criterion of Ref. \cite{Toschi}. PLM denotes the preformed local moment regime, SCR {the} regime of local moment screening, and FL stands for the Fermi liquid state. The critical interaction $U_{c2}$ of the MIT
taken from Ref. \cite{Triangular1} is indicated by the blue line, and irreducible vertex divergences \cite{Schafer} are shown by yellow and orange lines. The inset zooms the region near {the} MIT.
}
\label{fig:phase_diagram}
\end{figure}

The phase diagram, summarizing the above results, is shown in Fig. \ref{fig:phase_diagram}. The obtained boundary of the beginning of the formation of a local magnetic moment corresponding to the temperatures $T_{c,{\rm max}}$ of maxima of local charge compressibility 
is qualitatively similar to the interaction dependence of the local moment formation, obtained recently in Ref. \cite{Katsnelson}. The interaction dependence of the temperatures $T_{c,{\rm max}}$ repeats also qualitatively the line of the first divergence of the irreducible charge vertex, obtained previously in Ref. \cite{Toschi}. At first glance, this confirms {the} interpretation of vertex divergencies as a trace of local moment formation, proposed in Ref. \cite{ToschiAnd1}.
The temperatures $T_{c,{\rm max}}$ are however somewhat larger than the temperatures, at which first divergence of the irreducible charge $\Gamma_{\rm irr}$ vertex occurs, which may indicate that the local moment formation starts, in fact, earlier than the vertex $\Gamma_{\rm irr}$ diverges. Also, the first divergence line is characterized by odd in frequency eigenfunctions of the charge susceptibility $\chi_c^{\nu\nu'}$ \cite{Schafer}, while only even in frequency eigenfunctions contribute to local compressibility (see Refs. \cite{ToschiAnd1,ToschiInterplay,ToschiDoping,OurMott1} and the Supplemental Material \cite{SM}). 

The  temperatures $T_{c,{\rm min}}$, corresponding to the minima of $\chi_c(T)$, as we have discussed above, determine the complete formation of local magnetic moments, and separate the region of partially formed local moments (at $T>T_{c,{\rm min}}$) and their subsequent screening (at $T<T_{c,{\rm min}}$). The location of this boundary, as it is mentioned above, appears to be very close to the temperatures of the minima of the double occupancy (green dashed line with squares); 
the temperatures $T_{c,{\rm min}}$ are also sufficiently close to the {low-temperature} boundary of the plateau of $\mu_{\rm eff}^2$. {On the other hand, as we discuss in the Supplemental Material \cite{SM}, the temperature scale $T_{c,{\rm min}}$ is related to the half width of the central (quasiparticle) peak, which confirms that the screening of the local moment below $T_{c,{\rm min}}$ (and change of the temperature dependence of $\mu_{\rm eff}^2$ from the plateau to the Kondo behavior) occurs due to states at the quasiparticle peak of the spectral function.} 
We also note that the minima and maxima of local compressibility, {as well as Hubbard subbands of the spectral function}, are obtained only above
the interaction $U$, at which the first vertex divergence occurs (see also Ref. \cite{thesis}). With increasing interaction the line $T_{c,{\rm min}}$ approaches the {endpoint} of the critical interaction $U_{c2}$ of MIT,
where it joins with the crossover line between the bad metal and Mott insulator (not shown). 
This reminds us of a change of the critical exponent of resistivity $\rho\sim T^\beta$ from $\beta>2$ to $\beta<2$ at the boundary of a similar shape, located near the Widom (crossover from metal to insulator) line, discussed some time ago for frustrated magnetic systems \cite{resistivity}.  
The boundary between PLM and SCR regimes 
also qualitatively follows
the bendings of irreducible vertex divergence lines, obtained in Ref. \cite{Schafer}, which allows us to associate these bendings with the PLM-SCR crossover. 

At $T<T_K$ the Fermi liquid state of screened local moments appears; the interaction $U$ dependence of the Kondo temperature is shown in Fig. \ref{fig:phase_diagram}. For comparison, we also plot the results for the Kondo temperature from the ``fingerprint'' criterion, based on {a} comparison of $\chi_c^{\nu\nu'}$ at the lowest fermionic Matsubara frequencies \cite{Toschi,SM}. One can see that the two definitions of Kondo temperatures yield close results near MIT,
providing a ``universal" definition of the Kondo temperature in this regime. {In agreement with the results of Ref. \cite{HeldScal}, we obtain therefore two different energy scales near MIT, the Kondo temperature and $T_{c,{\rm min}}$. }However, with a decrease of the Coulomb interaction, the ``fingerprint'' criterion yields {an} overestimation of the Kondo temperature and turns into the boundary of the divergence of the irreducible vertex, obtained in Ref. \cite{Schafer}, above the temperature of the bending of {the} first divergence line. This shows that away from MIT
not only 
{the} lowest Matsubara frequencies 
contribute to screening, which reflects {the} widening of {the} central peak 
of the spectral function with decreasing interaction. 
It is plausible to assume that the  screened state is described by some linear combination of odd in frequency eigenfunctions of
the susceptibility $\chi_c^{\nu\nu'}$. 
This would be consistent with the fact that the local charge compressibility, which is contributed by even eigenfunctions of $\chi_c^{\nu\nu'}$, does not show any peculiarities at the Kondo temperature. 

In summary, we have studied the relation between spin and charge responses in different stages of local moment formation and screening. The formation of a local magnetic moment is signaled by {the} plateau of the temperature dependence of the effective magnetic moment $\mu^2_{\rm eff}=T\chi_s(0)$, the minimum of local charge susceptibility, and double occupation. With further reducing temperature the local moment is screened, the effective moment decreases, while local charge compressibility and double occupation increase. A strong increase of charge susceptibility versus {a} weak increase of double occupation demonstrates {the} importance of virtual transitions in local magnetic moment screening. Since local charge compressibility is affected by {the} formation of a local moment, we associate this process with {a} contribution of even in frequency eigenfunctions of 
the susceptibility 
$\chi_c^{\nu\nu'}$. Full screening of the local moment occurs at $T<T_K$. 
We show that in the vicinity of MIT,
$T_K$ is correctly described by the fingerprint criterion, while further away from the transition the latter criterion somewhat overestimates the Kondo temperature. 

In the present Letter, we neglect magnetic correlations due to the long-range order in the ground state. In this respect, the results are applicable to frustrated lattices and can be used to describe peculiarities of the spin liquid state \cite{SL,Triangular2,Triangular3}.  
More generally, the results of this Letter can be further used for {the} description of materials with almost formed local moments, such as Hund's metals, systems in the vicinity of MIT,
etc. {The relation of the obtained results to the recently pointed topological nature of MIT \cite{Mitchell} has to be further investigated.} Analytical studies of the relation of charge and spin responses in systems with local moments {are} also of certain interest.


The authors are grateful to A. Toschi and P. Chalupa for discussions and providing their data for the phase diagram of Ref. \cite{Toschi} and some unpublished data of Ref. \cite{thesis}. The authors acknowledge the financial support from the BASIS foundation (Grant No. 21-1-1-9-1) and the Ministry of Science and Higher Education of the Russian Federation (Agreement No. 075-15-2021-606). A. A. K. also acknowledges the financial support within the theme ``Quant" AAAA-A18-118020190095-4 of Ministry of Science and Higher Education of the Russian Federation.

\clearpage
\appendix
\renewcommand\theequation{A\arabic{equation}}
\renewcommand\thefigure{S\arabic{figure}}
\setcounter{equation}{0}
\setcounter{figure}{0}
\section*{Supplemental Material \\
\lowercase{to the paper }``L\lowercase{ocal magnetic moment formation and }K\lowercase{ondo screening in the half-filled single-band }H\lowercase{ubbard model}"}
\vspace{-0.4cm}
\centerline{T. B. Mazitov and A. A. Katanin}
\vspace{0.5cm} 
 
 \subsection{Eigenvalues of charge susceptibility}

The fermionic {frequency-resolved} local charge susceptibility is defined by
\begin{eqnarray}
\chi_c^{\nu \nu'}&=&T^2\sum_{\sigma\sigma'}\int_0^\beta d\tau_1 d\tau_2 d\tau_3\left[\langle T_\tau  c^\dagger_{i\sigma}(\tau_1) c^{}_{i\sigma}(\tau_2)\right.\notag\\ &\times&c^\dagger_{i\sigma'}(\tau_3) c^{}_{i\sigma'}(0)\rangle
-\langle T_\tau c^\dagger_{i\sigma}(\tau_1) c^{}_{i\sigma}(\tau_2) \rangle \notag\\ 
&\times&\left.\langle c^\dagger_{i\sigma'}(\tau_3) c^{}_{i\sigma'}(0)\rangle\right]e^{i\nu(\tau_1-\tau_2)+i\nu'\tau_3}.
\end{eqnarray}
where $T_\tau$ denotes the chronological ordering, $\nu,\nu'$ are the fermionic Matsubara frequencies, $i$ refers to the impurity site. We further define the (right) eigenvectors and eigenvalues of the charge susceptibility $\chi_c^{\nu\nu'}$ by 
\begin{equation}
    \sum_{\nu'}{ \chi_c^{\nu\nu'}\phi^R_{\nu'\alpha}=\lambda_\alpha \phi^R_{\nu\alpha}}.
\end{equation}
By introducing inverse (left) eigenvectors $\phi^{L}_{\nu\alpha}\equiv(\phi^{-1})_{\nu\alpha}$ 
(taken as a matrix inverse), we find
\begin{equation}
    \chi_c^{\nu\nu'}=\sum_\alpha \lambda_\alpha  \phi^R_{\nu\alpha}\phi^{L}_{\alpha\nu'}.
\end{equation}
Full local charge susceptibility $\chi_c=\sum_{\nu,\nu'}\chi_c^{\nu\nu'}$ is then expressed as $\chi_c=\sum_\alpha \chi_c^{\alpha}$, where
\begin{equation}
    \chi_c^\alpha=\lambda_\alpha \left(\sum_{\nu} \phi^R_{\nu\alpha}\right)\left(\sum_{\nu'}\phi^{L}_{\alpha\nu'}\right).\label{chialpha}
\end{equation}
According to Eq. (\ref{chialpha}) only eigenvalues, corresponding to the even in frequency eigenfunctions, contribute to charge susceptibility, cf. Ref. \cite{ToschiInterplay} of the paper. The odd in frequency eigenfunctions may, however, play important role in {the} description of screening of local magnetic moment, see Sect. 3 of this Material.

\begin{figure}[t]
    \includegraphics[width=0.9\linewidth]{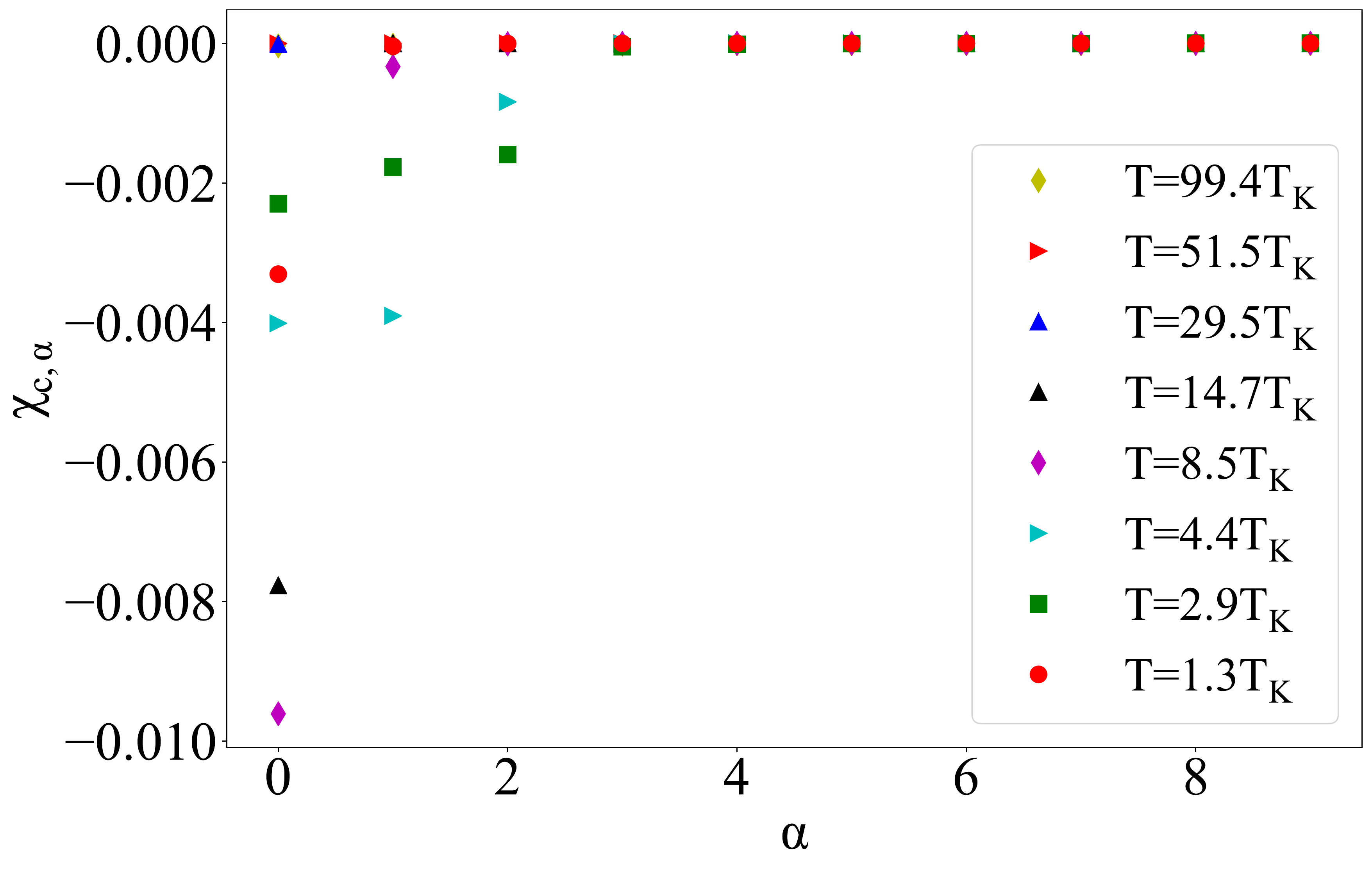}
    \includegraphics[width=0.9\linewidth]{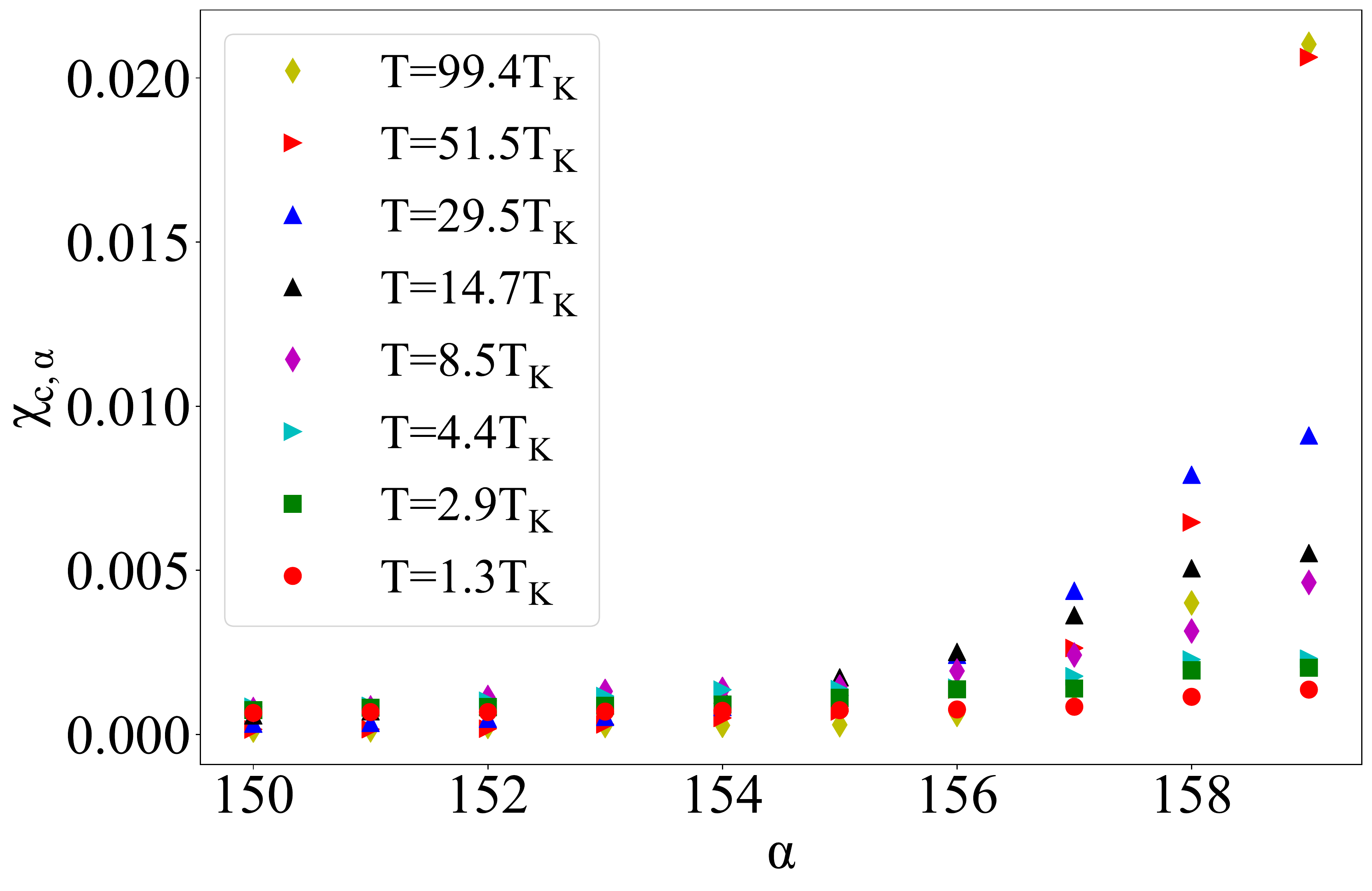}
\caption{The lowest (upper plot) and largest (lower plot) contributions to charge susceptibility 
{are} arranged in ascending order
at various values of temperature $T$ at the value of the Coulomb interaction $U = 2.0$. 
}
\label{fig:charge_susceptibility_components}
\end{figure}

One can see from Fig. \ref{fig:charge_susceptibility_components} that the largest contributions to charge susceptibility, which originate from positive eigenvalues, are suppressed with temperature. The subleading largest contributions are however non-monotonic at the temperatures near $T_{c,{\rm max}}$, providing maximum of charge susceptibility at $T=T_{c,{\rm max}}$. The non-monotonic behavior of charge susceptibility near $T=T_{c,{\rm min}}$ is entirely related to non-monotonic behavior of lowest (negative) contributions, which originate from negative eigenvalues (corresponding to even in frequency eigenfunctions) 
occurring due to passing irreducible vertex divergence lines.   

\subsection{Double occupation}

\begin{figure}[b]
\centerline{\includegraphics[width=1.0\linewidth]{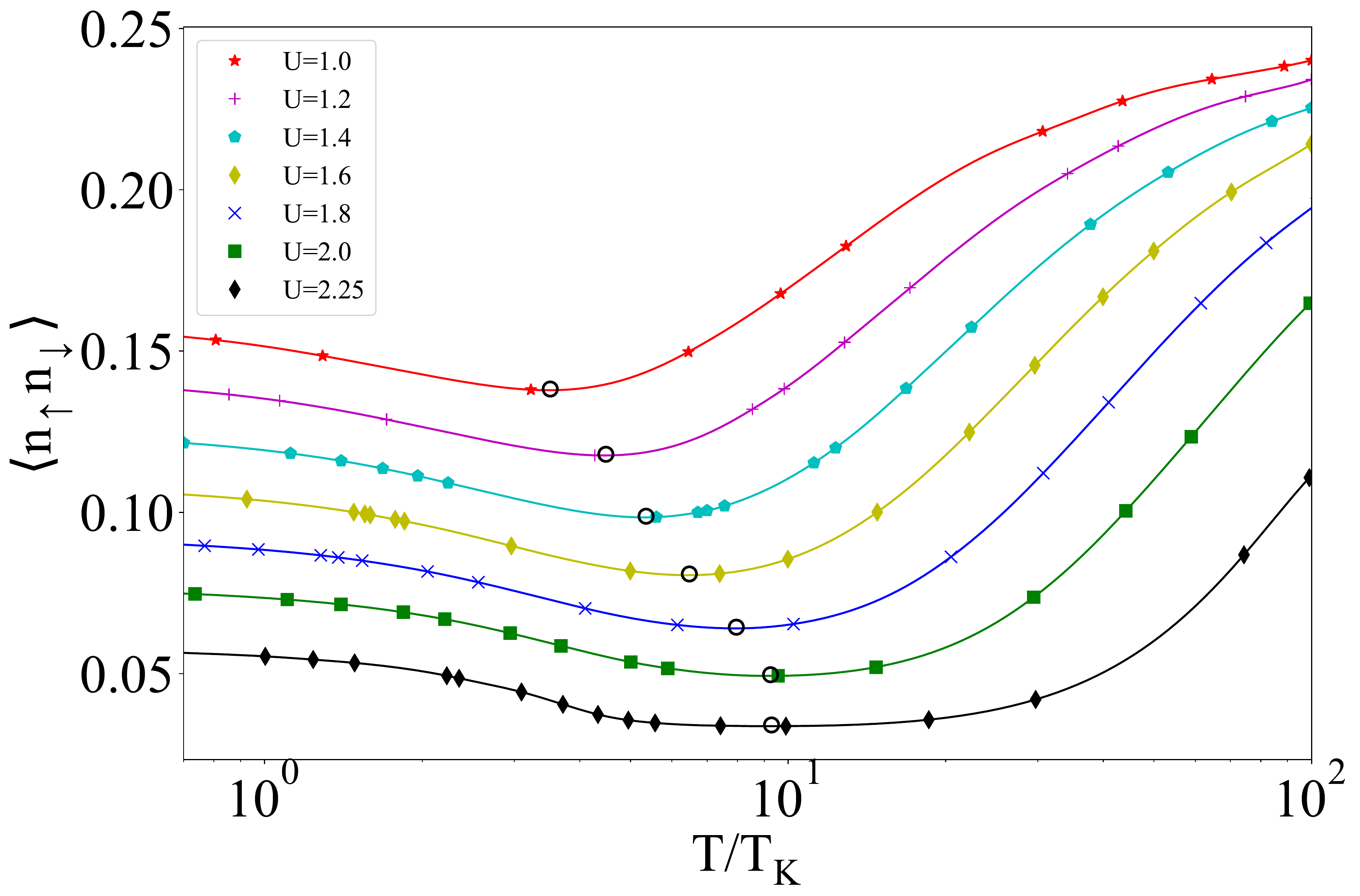}}
\caption{Temperature dependence of the double occupation for different values of the Coulomb interaction $U$. The open black circles indicate local minima of double occupation.}
\label{fig:nup_ndown}
\end{figure}

In Fig. \ref{fig:nup_ndown} we show temperature dependencies of double occupations $\langle n_{\uparrow} n_\downarrow \rangle$ for various $U$. As one can see, the double occupation decreases with decreasing temperature at $T\gtrsim 10T_K$. The minimum of $\langle n_\uparrow n_\downarrow \rangle$, which indicates the maximum of the square of the local moment, is observed at approximately the same temperatures $T_{c,{\rm min}}$, as extracted from the minima of local compressibility. {An} increase of double occupation with further decreasing temperature corresponds to {the} screening of local magnetic moments.

\subsection{Electron spectral functions}

\begin{figure}[h!]
\centerline{\includegraphics[width=0.95\linewidth]{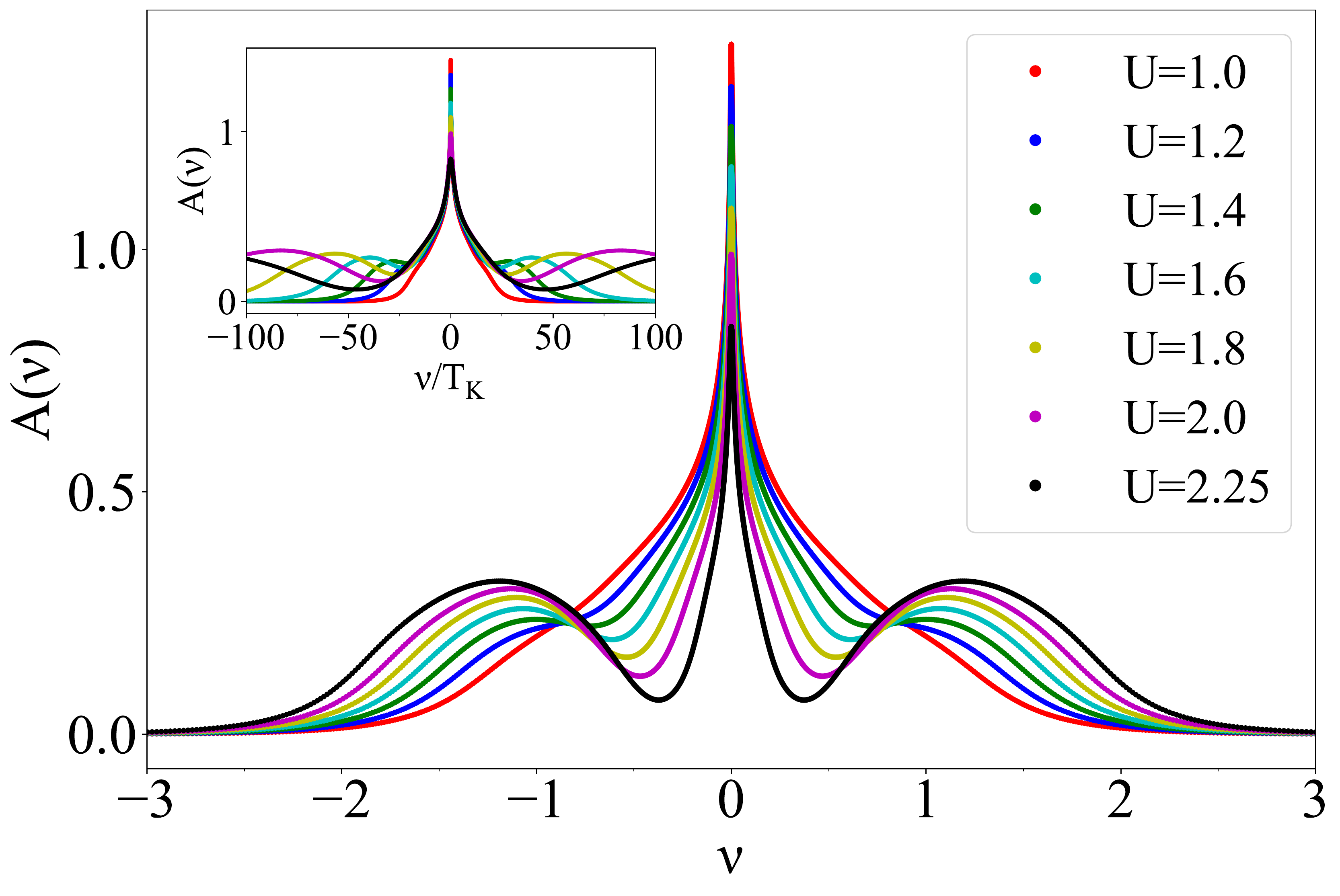}}
\caption{(Color online). The frequency dependence of the electron local spectral function $A(\nu)$ obtained by NRG approach at various values of the Coulomb interaction $U$ and temperature $T = 0.01$. Inset shows the plot of $A(\nu/T_K)$.}
\label{fig:nrg_freq_dep}
\end{figure}

\begin{figure}[t]
\centerline{\includegraphics[width=0.95\linewidth]{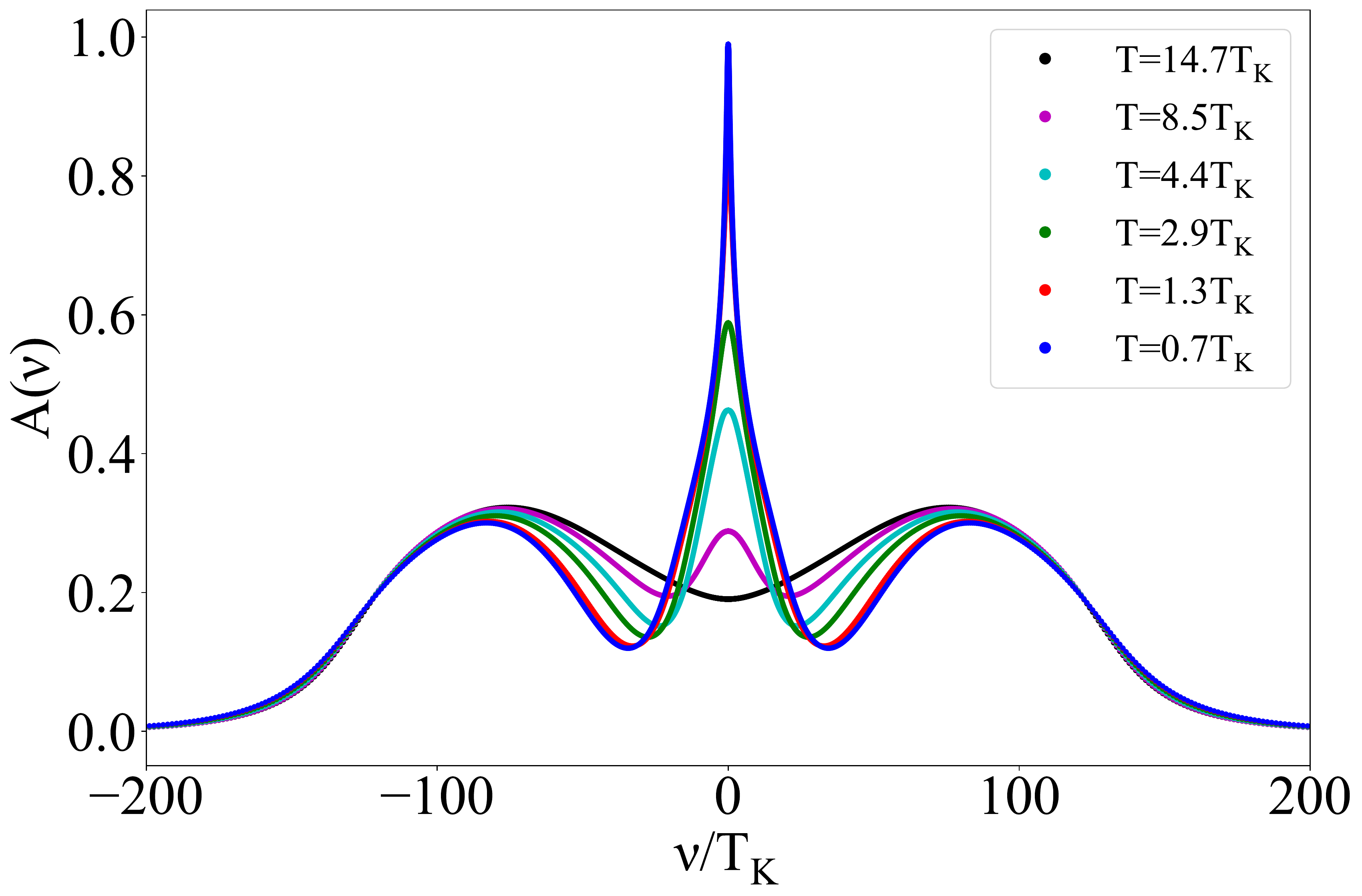}}
\caption{The frequency dependence of the spectral function at various values of the temperature $T$ and the Coulomb interaction $U = 2$, obtained by NRG approach.}
\label{fig:nrg_spectral_3}
\end{figure}

For completeness we also present in Figs. \ref{fig:nrg_freq_dep} and \ref{fig:nrg_spectral_3} the frequency dependence of the electron local spectral function $A(\nu)=(-1/\pi){\rm Im} G_{\rm loc}(\nu)$, where $G_{\rm loc}(\nu)$ is the local Green's function at the real frequency axis, obtained by NRG approach (we have verified that the analytical continuation of CT-QMC data produces close results). Fig. \ref{fig:nrg_freq_dep} shows the evolution of the spectral functions at low temperature with increasing interaction strength. One can see that Hubbard subbands appear already at $U=1.4$, on approaching the first divergence line in Fig. 4 of the paper. Although the quasi-particle peak width decreases with increase of $U$, in units of $T_K$ it changes only weakly with $U$ (see the inset). The peak remains much broader than $T_K$ in agreement with the results of Ref. \cite{HeldScal} of the paper, which suggested presence of two different energy scales ($T_K$ and the width of quasi-particle peak) near the MIT. Comparing the full half width $\nu_{\rm QP}$ of the central peak to the other temperature scales, discussed in the paper, we find that it is related to the temperature, $T_{c,{\rm min}}$ at which local moments are fully formed, by $\nu_{\rm QP}\simeq 5T_{c,{\rm min}}$. 

Fig. \ref{fig:nrg_spectral_3} shows temperature evolution of spectral functions  at the value of the Coulomb interaction $U=2.0$, when the local moments are formed at sufficiently low temperatures. Due to a peculiarity of the considered square lattice, which have a logarithmic divergent bare density of states, with reducing temperature the height of the peak of the spectral function increases due to small quasi-particle damping at temperatures of the order and below $T_K$ (i.e. in the Fermi liquid regime). {Although this can yield corrections to the universal dependence of the effective moment $\mu_{\rm eff}^2$ on $T$,
in agreement with Ref. \cite{Log} we do not find deviations from the Kondo model behavior at $T<T_K$ (see Fig. 1 of the paper). Apart from that, in units of the obtained $T_K$ the width of both, local spin susceptibility (see the inset of Fig. 2 of the paper) and spectral function (the inset of Fig. \ref{fig:nrg_freq_dep}) at low temperatures remain approximately constant, which shows correctness of the determination of $T_K$.} Also, as one can see from Fig. \ref{fig:nrg_spectral_3}, the maximal width of the peak remains approximately constant with decreasing temperature.

\subsection{The $2\times2$ fermionic frequency subspace and the ``fingerprint criterion"}
The ``fingerprint" criterion of Ref. \cite{Toschi} of the paper operates with $2\times 2$ subspace of Matsubara frequencies $\nu,\nu'=\pm \nu_1\equiv\pm \pi T$
and reads $\chi_c^{\nu_1 \nu_1}=\chi_c^{\nu_1, -\nu_1}$. As we argue in the main text, this criterion is applicable near Mott transition, {where in} the vicinity of the Kondo temperature $T\lesssim T_K$  one can restrict consideration by the {above-mentioned} subspace, which corresponds physically to the quite narrow width of central (quasiparticle) peak of the spectral function. On the other hand, at sufficiently high temperatures $T\gtrsim T_{c,{\rm max}}$ one can also restrict consideration to the abovementioned subspace since larger Matsubara frequencies give {an} irrelevant contribution in that regime, cf. Ref. \cite{Schafer} of the paper. 

In the above mentioned subspace we have 
\begin{eqnarray}
    \phi^{R}_{\nu,1}&=&\phi^{L}_{\nu,1}=(1,1)/\sqrt{2},\notag\\
    \phi^{R}_{\nu,2}&=&\phi^{L}_{\nu,2}=(1,-1)/\sqrt{2}
\end{eqnarray}
with eigenvalues $\lambda_{1,2}=\chi_c^{\nu_1,\nu_2}\pm\chi_c^{\nu_1,-\nu_1}$.
The charge susceptibility within the considered subspace $\chi_c=2\lambda_1$ is determined only by the eigenvalue $\lambda_1$. The eigenvectors $\phi^{R,L}_{\nu,1}$, which affect fermionic frequency summed charge susceptibility $\chi_c$ can be therefore related to the formation of {the} local moment. 
At the same time, the eigenvalue $\lambda_2$ vanishes when the ``fingerprint" criterion of Kondo screening is fulfilled. The corresponding eigenvectors $\phi^{R,L}_{\nu,2}$ can be therefore considered as responsible for the local moment screening near Mott transition. 
The antisymmetry of these vectors with respect to frequency 
may reflect the {so-called} orthogonality theorem \cite{orth}. Interestingly, the same antisymmetric vectors $\phi^{R,L}_{\nu,2}$ were recently used in Ref. \cite{OurMott1} of the paper for constructing Landau functional of Mott transition, which is also related to {the} disappearance of quasiparticle peak. 

In discussing the phase diagram of Fig. \ref{fig:phase_diagram} of the paper we conjecture that the conclusions from the considered $2\times 2$ frequency subspace on the parity of relevant eigenfunctions in various regimes remain valid in {a} broader range of $U$ in the strong coupling regime, where broader fermionic frequency range becomes important. 






\bibliographystyleApp{unsrt}


\begin{thebibliography}{1}
\bibitem{Notew} {At large frequencies the real part of susceptibility is negative in accordance with the asymptotic form $\omega^2 {\rm Re}\chi_s(\omega)\rightarrow -(2/\pi)\int_0^\infty d\omega' \omega' {\rm Im} \chi_s(\omega')$ at $|\omega|\rightarrow \infty$.}
\end{thebibliography}
\end{document}